\documentclass[11pt]{article}
\usepackage[]{acl}
\usepackage{lmodern}
\usepackage[T1]{fontenc}
\usepackage[utf8]{inputenc}
\usepackage{latexsym}
\usepackage{booktabs}
\usepackage{graphicx}
\usepackage{amsmath}
\usepackage{amssymb}
\usepackage{xcolor}
\makeatletter
\newcommand{\FloatBarrier}{\par\ifx\@deferlist\@empty\else\clearpage\fi}
\makeatother
\PassOptionsToPackage{hyphens}{url}
\usepackage{url}
\title{MoE Pretraining:\\ Parallelism Layout Changes the Model, Not Quality}
\author{
  Esther Xin \\
  Independent Researcher \\
  \texttt{yixin@microsoft.com} \\
  {\small Code and data: \url{https://github.com/ethxin0011/moe_parallelism}}
}

\begin{document}
\maketitle

% =====================================================================
\begin{abstract}
Large Mixture-of-Experts (MoE) models train across many machines, and the way work is
split---the \emph{parallelism layout}---changes from run to run. You might resume a job
on a different GPU count, resize a cluster, or run an ablation on a new setup.In exact arithmetic the layout should not matter. In real arithmetic it changes the order in
which numbers are added, and some MoE rules (load balancing, expert capacity) are often computed
per machine. We ask a simple question: \emph{does the layout change the trained MoE model?} We
train the same small MoE language models (25M--117M parameters, FineWeb-Edu, 20M tokens) with
the same seed, data and hyperparameters under 9 data-parallel/expert-parallel layouts, in fp32 on
CPUs and fp16 on a GPU, and compare every run against how much a \emph{new random seed} changes
the model. Across 80 training runs we find: (1) the layout does not change model quality, but
it does change the model---by the end of training the predictions differ about 80\% as much as
a new seed would make them, and over 90\% of routing decisions differ; any layout, a one-bit
change to one weight, or a different CPU thread count has the same effect. (2) The cause is the
router: routing decisions that flip between runs are almost exactly those where two experts
were nearly tied (AUC 0.97--1.00 in the first 16 training steps). (3) Splitting experts more finely makes
near-ties more common (2.7\% of decisions with 8 experts, 23\% with 64) and routing less
reproducible (68\% vs.\ 100\% of decisions differ), but does \emph{not} make the outputs drift
further. (4) A dense model of the same active size is far more stable: in fp32 it ends 0.12$\times$
as far apart as a new seed, and in fp16 0.58$\times$, while the MoE reaches about 0.8$\times$ at
both precisions and gets there about twice as fast. (5) When a pretraining run is resumed on a
new layout, the timing matters: switching at 84\% of pretraining (step 1{,}024 of 1{,}221) barely
changes the final model, while switching at 10\% (step 128, just after learning-rate warm-up)
changes it substantially. We will release the code, all run
logs, and the measured data.
\end{abstract}

% =====================================================================
\section{Introduction}
\label{sec:intro}

Mixture-of-Experts (MoE) models replace one large feed-forward block with many smaller
``experts'' and a small \emph{router} that sends each token to a few of them
\citep{shazeer2017moe,lepikhin2021gshard,fedus2022switch}. Most frontier open models now use
this design, often with many fine-grained experts \citep{dai2024deepseekmoe,yang2025qwen3}.
Training them needs many machines, and the work is split in two main ways: \emph{data
parallelism} (DP), where each machine processes part of the batch, and \emph{expert
parallelism} (EP), where each machine holds part of the experts and tokens are exchanged
between machines \citep{lepikhin2021gshard,yan2026megatronmoe}. We call the choice of DP and
EP sizes the \emph{layout}.

In practice the layout is not fixed. A job is preempted and resumed on fewer GPUs, a cluster
grows, a team re-runs an ablation on a different machine count. The usual assumption is that
this is harmless: with the same seed, data and global batch, the mathematics is identical. But
computers add numbers in a fixed number of digits, so $(a+b)+c$ and $a+(b+c)$ can differ in the
last digit, and changing the layout changes the order in which gradients and expert outputs are
added up. On top of that, several MoE rules are commonly computed \emph{per machine} rather than
over the whole batch---the load-balancing statistics \citep{qiu2025demons}, the capacity limit
that decides which tokens are dropped, and the counts used by loss-free balancing
\citep{wang2024lossfree}. For these, a new layout changes what is computed, not just its
rounding.

\paragraph{Our question.} \emph{Does the parallelism layout change the trained MoE model---and if
so, why, and does it get worse as experts get more fine-grained?} We answer it with a
controlled study. We train the same models with the same seed, data order and global batch
under different layouts, using real multi-process training with real communication between
processes, and compare every pair of runs against three yardsticks: a new random seed, a new
data order, and a change of a single weight by the smallest possible amount (one unit in the
last place, ``1 ULP''). We separate two kinds of layout effects: \emph{numerical} (only the
order of additions changes) and \emph{definitional} (a per-machine rule switched on). We then
look inside the router to find what turns tiny rounding differences into different models, and
we repeat the study with 8 to 64 experts, at three model sizes, at two precisions, and with a
dense model of the same active size.

\paragraph{Main findings.}
\begin{enumerate}
\item \textbf{Same quality, different model.} Every layout ends with the same validation loss
as the reference, within seed-to-seed variation. But the predictions end about
\textbf{80\% as different as a new seed would make them}, and over 90\% of routing decisions
differ. All eight layouts, a 1-ULP change to one weight, and a different CPU thread count are
indistinguishable from one another (\S\ref{sec:res-main}).
\item \textbf{Per-machine rules start large but end the same.} Computing load-balancing or
capacity per machine creates differences about 400$\times$ larger at the start of training, yet
the final difference is the same as for a pure layout change (\S\ref{sec:res-semantic}).
\item \textbf{The cause is near-ties in the router.} The routing decisions that flip between two
runs are almost exactly the ones where the router's last chosen expert and the best unchosen
expert had nearly equal scores. Their score gap is several hundred times smaller than usual,
and they flip 65--130$\times$ more often than other decisions (\S\ref{sec:res-mechanism}).
\item \textbf{More experts: less reproducible routing, same output drift.} With total and active
size fixed, going from 8 to 64 experts raises the share of near-tie decisions from 2.7\% to
23.3\% and the share of routing decisions that differ at the end from 68\% to 100\%. The
difference in the model's outputs, however, stays at about 0.8$\times$ the seed level for every
expert count (\S\ref{sec:res-granularity}).
\item \textbf{MoE vs.\ dense, and the role of precision.} A dense model of the same active size
barely moves in fp32 (0.12$\times$ the seed level, and a straight line between the two models
has no loss bump). In fp16 it moves more (0.58$\times$). The MoE reaches about 0.8$\times$ at
both precisions, and reaches the 0.01 level of difference at step 64 instead of 128
(\S\ref{sec:res-dense}).
\item \textbf{Layout sensitivity is pronounced in early pretraining but muted in late pretraining.} In the early phase,
resuming on a new layout at 10\% of pretraining (step 128, shortly after learning-rate
warm-up, while the loss is still dropping fast) changes the final model substantially.
In the late phase, resuming at 84\% of pretraining (step 1{,}024 of 1{,}221, after the loss
has fallen from 9.8 to about 4.7) leaves the final model almost unchanged. In both phases,
a layout switch behaves exactly like a 1-ULP nudge (\S\ref{sec:res-resume}). All experiments
concern pretraining from scratch: ``early'' and ``late'' denote phases within pretraining,
not separate mid-training or post-training stages, which we did not study.
\end{enumerate}

\paragraph{Why it matters.} If you only care about loss, layouts are interchangeable. If you
care about getting \emph{the same model} back---to debug a loss spike, to compare two methods
fairly, to reuse routing decisions in reinforcement learning \citep{ma2025r3}, or to prune or
interpret specific experts---then for an MoE the layout is not a detail. Higher precision alone
does not fix this, and finer-grained experts make the routing part worse.

% =====================================================================
\section{Background in Plain Terms}
\label{sec:background}

\paragraph{Router and near-ties.} For every token, the router gives each expert a score and
keeps the top $k$. If the $k$-th best and the $(k{+}1)$-th best scores are almost equal---a
\emph{near-tie}---a tiny rounding difference can swap which expert gets the token. That expert
then receives a different gradient, and the two runs are no longer on the same path. We measure
the \emph{margin} of a decision as the gap between the $k$-th and $(k{+}1)$-th score, divided by
the spread of that token's scores, and call a decision a near-tie when the margin is below 0.01.

\paragraph{Layouts.} We write a layout as (DP, EP). With EP $=2$, for example, each group of two
processes shares the experts half-and-half and exchanges tokens inside the group. We keep the
\emph{global} batch the same in every layout, so each step sees exactly the same 32 sequences in
the same order.

\paragraph{Two kinds of layout effects.} A \emph{numerical} effect changes only the order of
additions. A \emph{definitional} effect changes what is computed: for example, the
load-balancing loss uses expert usage counted on one machine instead of the whole batch
\citep{qiu2025demons}, or the capacity limit is applied per machine. In our code every such
rule can be set to ``global'' (the same in every layout) or ``per-rank'' (the common
implementation), so we can switch each one on separately.

\paragraph{How we measure ``different''.} We compare two runs at the same step on the same
16{,}384 held-out tokens. Our main measure is the \textbf{KL divergence} between their predicted
next-token distributions: 0 means identical predictions. We also report how often their top
prediction agrees, how many routing decisions differ, and the \textbf{loss barrier}: the largest
rise in loss along the straight line between the two sets of weights
\citep{frankle2020lmc}. A barrier near zero means the two models sit in the same valley of the
loss landscape. To make numbers easy to read we divide each KL by the KL between two runs that
differ only in their random seed (``vs.\ own seed''). A value of 1.0 means ``as different as a
new seed''.

% =====================================================================
\section{Related Work}
\label{sec:related}

\paragraph{Training is sensitive to tiny changes.} \citet{summers2021nondeterminism} showed that
all sources of training randomness have a similar effect on how different the resulting models
are, and that changing one bit of the initial weights is enough to produce a very different
model. \citet{kwok2025butterfly} showed that tiny perturbations make runs drift apart mainly
during an initial phase of training that they describe as chaotic, and that this sensitivity
fades as training proceeds.
\citet{zhuang2022randomness} measured how tooling and hardware choices add randomness. These
works study dense models and generic noise. We study the parallelism layout specifically,
separate its numerical and definitional parts, compare MoE against dense directly, and explain
the MoE case through the router.

\paragraph{Layout-invariant computation.} \citet{he2025defeating} showed that run-to-run
differences in LLM inference come from batch-dependent addition order and built kernels that
remove them. \citet{zhang2026tbik} made inference and training produce bit-identical results
across tensor-parallel sizes. NVIDIA's Megatron Core supports bitwise-deterministic pretraining,
which guarantees identical results only when, among other things, the parallel configuration is
kept fixed \citep{nvidia2026determinism}. EasyScale keeps accuracy consistent when the number of
GPUs changes by preserving data-parallel behaviour exactly \citep{li2023easyscale}. These works
remove the difference; we measure how large it is when it is not removed, and why MoE models
are hit harder.

\paragraph{MoE routing instability.} In MoE reinforcement learning, small numerical differences
between the inference and training engines change which experts are chosen, which can make
training collapse; replaying the inference engine's routing fixes this \citep{ma2025r3}. We show
the same mechanism during pretraining, caused by the layout, and we identify which decisions flip.

\paragraph{Per-machine load balancing.} \citet{qiu2025demons} showed that computing the
load-balancing statistics over the global batch instead of each micro-batch improves quality and
expert specialization in large models. Loss-free balancing adjusts a per-expert bias from expert
counts \citep{wang2024lossfree,jin2026idbalancing}. These works ask which rule gives a better
model. We ask how much switching the rule between per-machine and global changes the model, and
how that compares with the pure numerical effect.

% =====================================================================
\section{Experimental Setup}
\label{sec:setup}

\paragraph{Models and data.} We train decoder-only language models from scratch in the style of
Qwen3-MoE \citep{yang2025qwen3}: QK-Norm, no bias in the attention projections, fine-grained
experts, top-$k$ routing and no shared expert (Table~\ref{tab:setup}). The base MoE has about 25M
parameters with about 10M active per token. The dense control has one feed-forward block with
the same active size. We train on FineWeb-Edu \citep{penedo2024fineweb} with a 16k-token
vocabulary, a global batch of 32 sequences of 512 tokens, and 1{,}221 steps (20M tokens). The
learning rate is constant after warm-up, so every run sees the same schedule.

\paragraph{Real parallel training.} Every multi-process run uses real processes that communicate
through PyTorch's gloo backend: gradients are summed across processes, and with EP $>1$ tokens and
expert outputs are exchanged between processes. CPU runs use fp32 on Azure E16s\_v3 nodes, with
the math libraries pinned to one instruction set so that the CPU generation is not a hidden
factor. GPU runs use fp16 with loss scaling on a single NVIDIA T4, with up to four processes
sharing the GPU. Before any training, a correctness check confirmed that every layout computes
the same gradients as a single process (relative error below $10^{-4}$ in fp32), that two runs of
the same layout are bit-identical, and that checkpoints move between layouts without change
(Appendix~\ref{app:checks}). So any difference we see between layouts is caused by the layout,
not by a bug or by random scheduling.

\paragraph{Runs.} Table~\ref{tab:experiments} lists the 80 training runs. The reference run uses
one process. Each other run changes exactly one thing.

\begin{table}[!htbp]
\centering
\includegraphics[width=\linewidth]{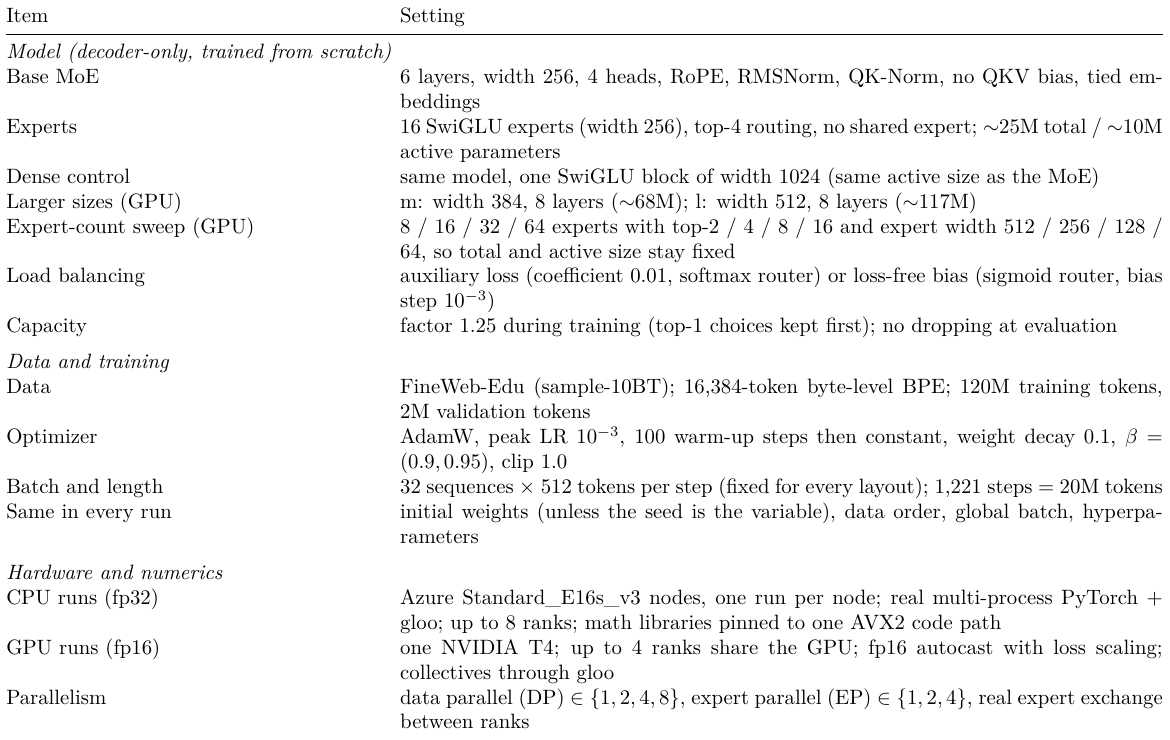}
\caption{Model, data, training and hardware settings.}
\label{tab:setup}
\end{table}

\begin{table}[!htbp]
\centering
\includegraphics[width=\linewidth]{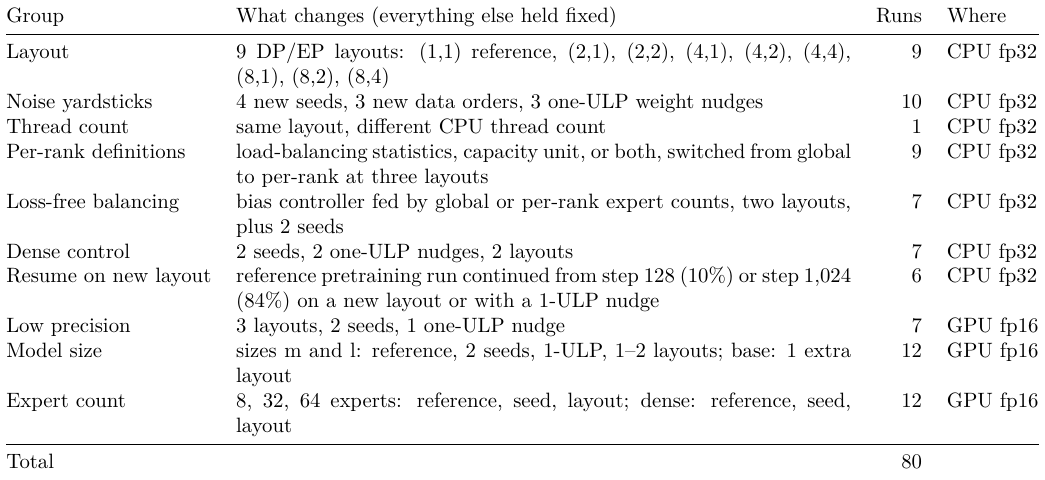}
\caption{The 80 training runs. Each run differs from its reference in one thing only.}
\label{tab:experiments}
\end{table}

% =====================================================================
\section{Findings}
\label{sec:results}

\FloatBarrier
\subsection{The layout changes the MoE model, but not its quality}
\label{sec:res-main}

\begin{figure}[!htbp]
\centering
\includegraphics[width=\linewidth]{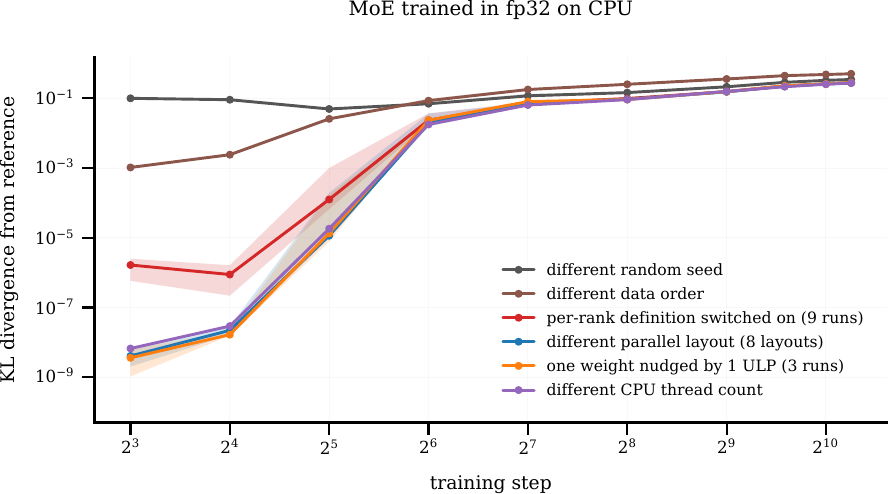}
\caption{Difference from the reference run during training (MoE, fp32 on CPU). Lines are
medians; shaded areas cover all runs in a group. Layout changes, 1-ULP nudges and a thread-count
change start about $10^{-9}$ and end at about 80\% of the seed level. Switching on a per-machine
rule starts about 400$\times$ higher but ends in the same place.}
\label{fig:curves}
\end{figure}

Figure~\ref{fig:curves} follows the difference from the reference over training, and
Table~\ref{tab:final} gives the final numbers. Three things stand out.

\textbf{It starts tiny and grows fast.} At step 8, a layout change has made the predictions
differ by a KL of only about $4\times10^{-9}$, roughly 25 million times less than a new seed.
By step 64 the difference has grown to about $2\times10^{-2}$, and by the end of training it is
0.280---\textbf{0.80$\times$ the seed level}. The top prediction agrees on only 69\% of tokens
(66\% for a new seed), and 92\% of routing decisions differ.

\textbf{Every small change looks the same.} The eight layouts, the three 1-ULP nudges and the
different CPU thread count follow one curve from step 32 onward and end between 0.277 and 0.283.
Training in fp16 on the GPU instead of fp32 on the CPU also ends at 0.279. The particular layout
does not matter: \emph{any} change to the arithmetic is enough.

\textbf{Quality does not change.} The final validation loss of the reference and the eight
layouts ranges from 4.576 to 4.610, inside the range of the four new-seed runs (4.582--4.626;
Appendix Table~\ref{tab:quality}). The layout runs are also much closer to the reference than a
new seed in weight space: the loss barrier between them is about 0.42 nats, against about 3.5 for
a new seed. They are different models that share most of their structure.

For comparison, a new data order makes the model more different than a new seed does at this
budget (0.51 vs.\ 0.35), so the order of the data is a larger source of variation than the
layout.

\begin{table}[!htbp]
\centering
\includegraphics[width=\linewidth]{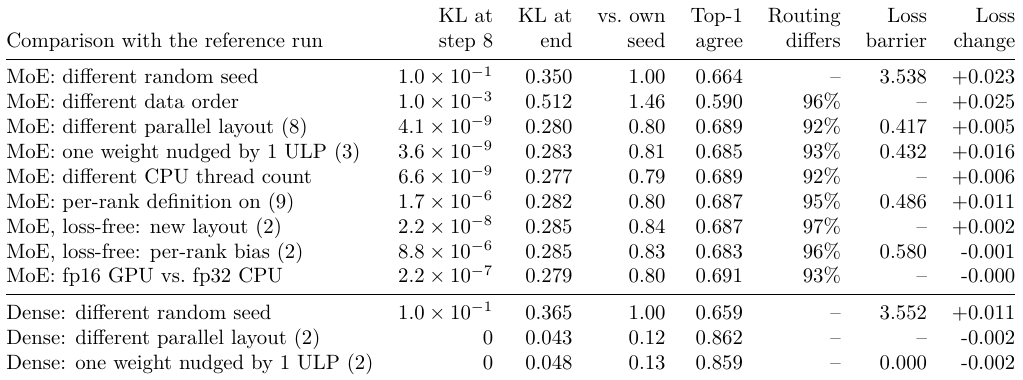}
\caption{Difference from the reference at step 8 and at the end of training. ``vs.\ own seed''
divides the final KL by the KL between two seeds of the same model type. ``Routing differs'' is
the share of (token, layer) decisions whose chosen expert set differs; it is not shown for seed
runs, where expert numbers are not comparable. ``Loss change'' is the final validation loss minus
the reference's. Counts in brackets are the number of runs averaged.}
\label{tab:final}
\end{table}

\FloatBarrier
\subsection{Per-machine rules start large but end the same}
\label{sec:res-semantic}

Switching the load-balancing statistics or the capacity unit from global to per-machine is a
real change to the training rule, not just to rounding. At step 8 it makes the model differ
about 400$\times$ more than a pure layout change ($1.7\times10^{-6}$ vs.\ $4\times10^{-9}$; red
band in Figure~\ref{fig:curves}). But by step 64 the two are indistinguishable, and the final
difference is the same (0.282 vs.\ 0.280, compared against the same layout without the
per-machine rule). The same holds for the loss-free bias controller fed by per-machine counts
(0.285). None of these rules changed the final loss beyond seed variation at our scale. The one
visible side effect is that per-machine capacity at 8 processes drops more tokens (1.8\% vs.\
1.2\%). This does not mean the per-machine rules are harmless at large scale, where global
load balancing is known to improve quality \citep{qiu2025demons}; it means that, for the
question ``do I get the same model back'', a definitional change and a rounding change end up in
the same place.

\FloatBarrier
\subsection{Why: near-ties in the router}
\label{sec:res-mechanism}

\begin{figure}[!htbp]
\centering
\includegraphics[width=\linewidth]{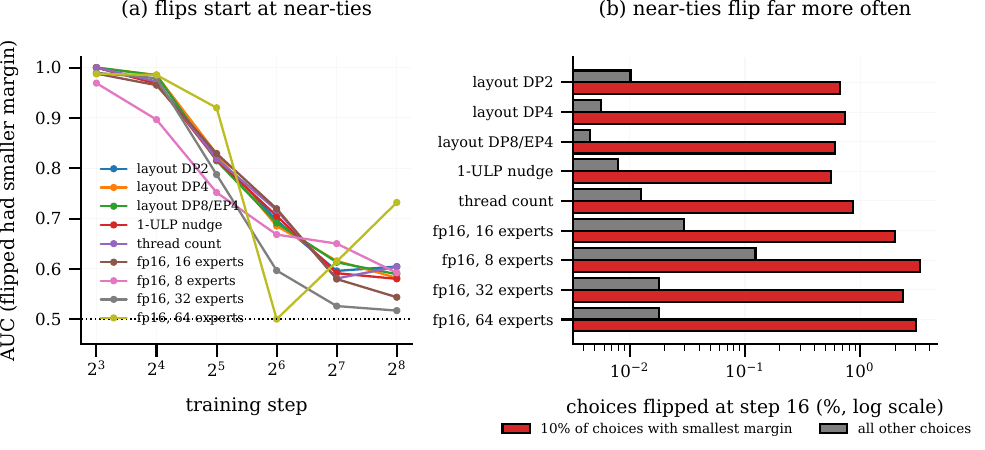}
\caption{(a) Probability that a flipped routing decision had a smaller margin than an unchanged
one (0.5 = no relation). Up to step 32, flips happen almost only at near-ties. (b) At step 16, the 10\% of
decisions with the smallest margins flip far more often than the rest (log scale).}
\label{fig:mechanism}
\end{figure}

To find what turns rounding differences into different models, we look at every routing
decision at every saved step and ask which ones flipped between the reference and the changed
run (Figure~\ref{fig:mechanism}, Table~\ref{tab:mechanism}).

\textbf{Flips happen at near-ties.} In the first 16 training steps, the decisions that flip are almost exactly
the ones with the smallest margin. If we pick a random flipped decision and a random unchanged
one, the flipped one has the smaller margin 97--100\% of the time (AUC 0.97--1.00 at step 8, for
every layout, the 1-ULP nudge and the thread change). At step 16, the median margin of a flipped
decision is 380--700$\times$ smaller than that of an unchanged one, and the 10\% of decisions
with the smallest margins flip 65--130$\times$ more often than the rest.

\textbf{After that, everything moves.} Once enough decisions have flipped, the weights
themselves start to differ, and flips spread to decisions with large margins too. From step 128
on, the AUC falls to about 0.6 and most decisions differ. Near-ties explain where the divergence
\emph{starts}; once started, it spreads on its own.

\textbf{Per-machine rules flip decisions directly.} With a per-machine capacity or balancing
rule, 2.5--3.4\% of decisions already differ at step 16 (vs.\ 0.06--0.10\% for a pure layout
change), and the margin explains them less well (AUC about 0.95), because the rule itself
chooses differently, not only at near-ties.

\begin{table}[!htbp]
\centering
\includegraphics[width=\linewidth]{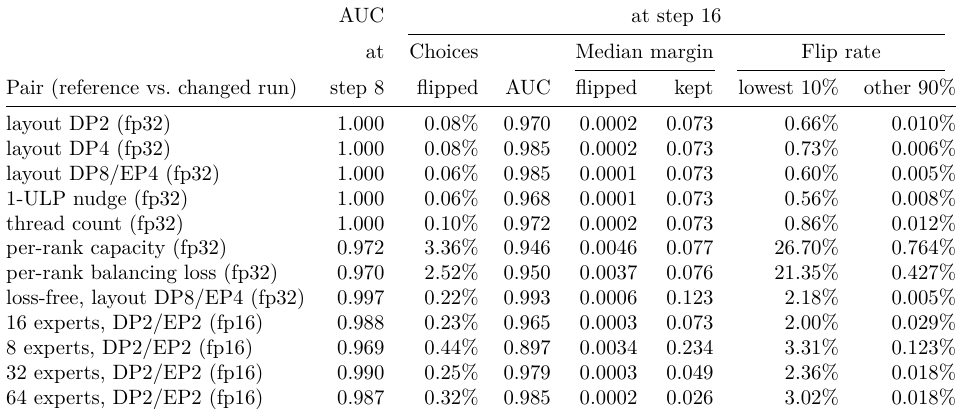}
\caption{Routing flips and margins at training steps 8 and 16 (0.7\% and 1.3\% of pretraining). ``AUC'' is the probability that a flipped
decision had a smaller margin than an unchanged one. Margins are divided by the spread of each
token's router scores.}
\label{tab:mechanism}
\end{table}

\FloatBarrier
\subsection{More experts: routing gets less reproducible, outputs do not drift further}
\label{sec:res-granularity}

\begin{figure}[!htbp]
\centering
\includegraphics[width=\linewidth]{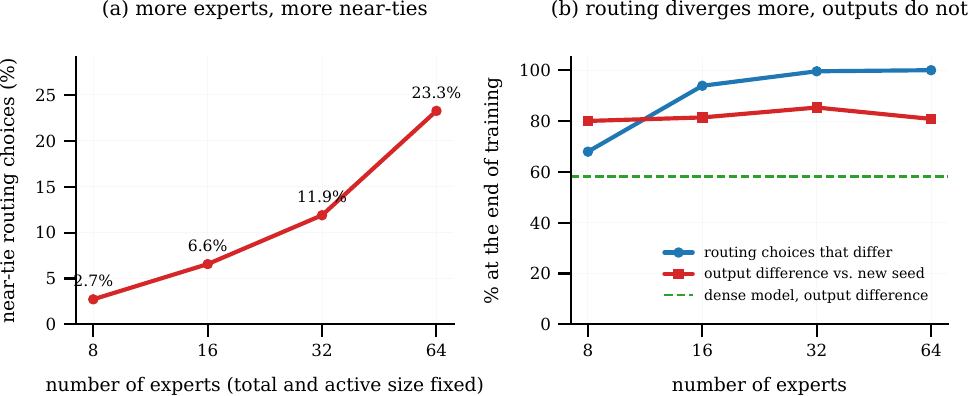}
\caption{Expert count with total and active size fixed (fp16, GPU). (a) Share of routing
decisions that are near-ties, averaged over steps 8, 16 and 32. (b) After a layout change: share of routing decisions that differ
at the end, and the output difference relative to a new seed. The dashed line is the dense
model.}
\label{fig:granularity}
\end{figure}

Modern MoEs split the feed-forward capacity into many small experts
\citep{dai2024deepseekmoe,yang2025qwen3}. If near-ties drive the effect, finer experts should
make it stronger. We test this on the GPU with 8, 16, 32 and 64 experts, choosing top-$k$ and
expert width so that total and active size stay exactly the same (Figure~\ref{fig:granularity},
Table~\ref{tab:granularity}).

\textbf{Near-ties roughly double each time the number of experts doubles}: 2.7\%, 6.6\%,
11.9\% and 23.3\% of routing decisions at steps 8--32 for 8, 16, 32 and 64 experts. With more experts,
more of them compete for each slot, and their scores sit closer together.

\textbf{Routing becomes less reproducible.} At the end of training, a layout change leaves 68\%
of routing decisions different with 8 experts, 94\% with 16, and essentially all of them with 32
or 64.

\textbf{But the outputs do not drift further.} The final difference in predictions is 0.80,
0.81, 0.85 and 0.81 of the seed level for 8 to 64 experts, and every expert count reaches the
0.01 level at the same step (64). Many small experts appear to be largely interchangeable: a token
sent to a different expert still gets a similar result. For practitioners this cuts both ways:
fine-grained MoEs are no worse in output stability, but their expert assignments are almost
impossible to reproduce across layouts, which matters for routing replay, expert pruning and
interpretability.

\begin{table}[!htbp]
\centering
\includegraphics[width=\linewidth]{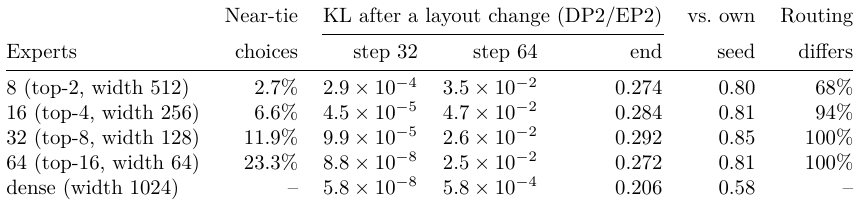}
\caption{Effect of the number of experts (total and active size fixed; fp16 on one T4). The layout
change is DP2/EP2 against the single-process reference. ``Near-tie choices'' is the mean over the
first three saved steps.}
\label{tab:granularity}
\end{table}

\FloatBarrier
\subsection{MoE versus dense, and the role of precision}
\label{sec:res-dense}

\begin{figure}[!htbp]
\centering
\includegraphics[width=\linewidth]{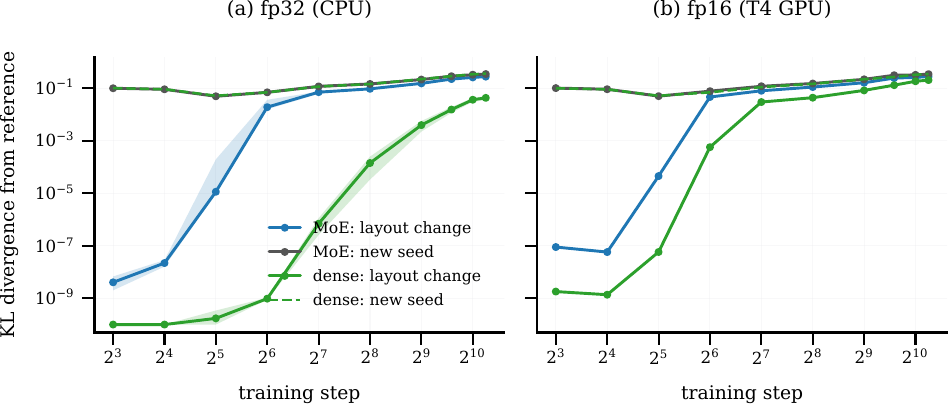}
\caption{Difference after a layout change, MoE vs.\ a dense model of the same active size. (a)
fp32 on CPU: the dense model stays almost identical; the MoE diverges. (b) fp16 on GPU: both
diverge, but the MoE does so earlier and further.}
\label{fig:dense}
\end{figure}

Figure~\ref{fig:dense} compares the MoE with a dense model of the same active size.

\textbf{In fp32, the dense model barely moves.} Its predictions match the reference to
within what we can measure (KL below $10^{-10}$) for the first 16 steps after a layout change, are still only
$7\times10^{-7}$ at step 128 (the MoE is at $7\times10^{-2}$), and end at 0.043, or 0.12$\times$
its seed level. The straight line between the dense reference and its 1-ULP run has no loss
bump at all: they are the same solution. The MoE, in contrast, ends at 0.80$\times$ with a loss
barrier of 0.42.

\textbf{In fp16, the dense model moves too, but more slowly and less.} With coarser arithmetic, the
dense model's layout difference ends at 0.58$\times$ its seed level, against 0.81$\times$ for the
16-expert MoE. The MoE still gets there much faster: every MoE crosses the 0.01 level at step 64,
the dense model at step 128, and at step 64 the MoEs differ 40--80$\times$ more than the dense
model.

\textbf{What this says.} For an MoE, the size of the rounding noise hardly matters: a difference
in the last bit is enough to flip a near-tie, and once one decision flips, the run diverges. The
MoE ends at about 80\% of the seed level whether we use fp32 or fp16. For a dense model, the
outcome depends on how large the rounding noise is: small noise in fp32 stays small, larger noise
in fp16 grows. In other words, the router makes MoE divergence close to all-or-nothing.

\FloatBarrier
\subsection{Resuming on a new layout is safe in the late phase, not the early phase}
\label{sec:res-resume}

\begin{figure}[!htbp]
\centering
\includegraphics[width=\linewidth]{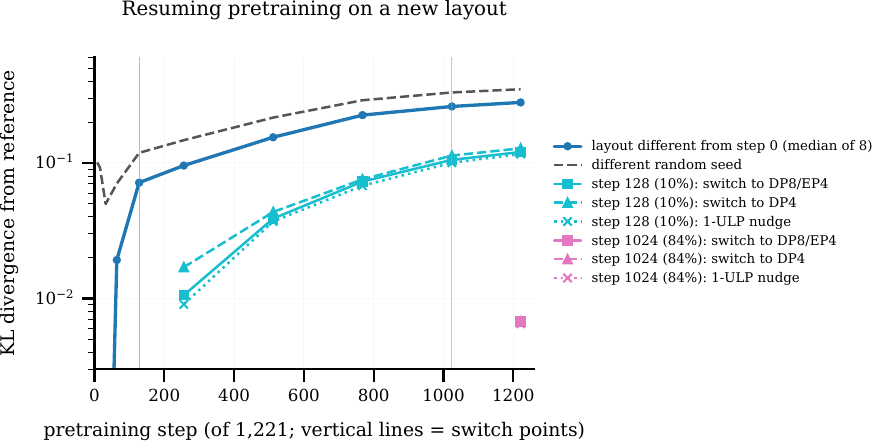}
\caption{Continuing the reference pretraining run on a different layout at step 128 (10\% of
pretraining) or step 1{,}024 (84\%), compared with a 1-ULP nudge at the same point and with runs
whose layout differs from step 0.}
\label{fig:resume}
\end{figure}

Elastic training resumes a job on whatever machines are available. To test how timing affects
this, we resumed the reference pretraining run from two checkpoints on two other layouts, as
well as with a 1-ULP nudge (Figure~\ref{fig:resume}):
\begin{itemize}
\item \textbf{Early phase (10\% of pretraining;} step 128 of 1{,}221; 2.1M of 20M tokens). This
occurs just 28 steps after the 100-step learning-rate warm-up ends, while validation loss is
still dropping sharply (6.8 at step 100 to 6.1 at step 200).
\item \textbf{Late phase (84\% of pretraining;} step 1{,}024; 16.8M tokens). By this point, the
learning rate has stayed constant for over 900 steps and validation loss has nearly flattened
at 4.70 (step 1{,}000), close to the final 4.58.
\end{itemize}
Switching at 84\% leaves the final model almost unchanged (KL 0.007, about 2\% of the seed level;
15\% of routing decisions differ). Switching at 10\% changes it much more (KL 0.12--0.13, about
35\% of the seed level; 58--61\% of routing decisions differ). In both cases the loss barrier is
zero, so the resumed model stays in the same valley as the original, unlike a run whose layout
differs from step 0 (barrier 0.42). And in both cases a layout switch behaves exactly like a 1-ULP
nudge. This matches the observation that training is most sensitive in its initial phase
\citep{kwok2025butterfly}. For MoE pretraining, our two checkpoints show that this sensitive period
still has a large effect 10\% into the run and a small one at 84\%. With only two switch points we
cannot say exactly where in between the effect becomes small, and all our runs use a constant
learning rate after warm-up, so we do not know how a switch during learning-rate decay,
mid-training or post-training would behave.

\FloatBarrier
\subsection{Model size}
\label{sec:res-size}

At three sizes (25M, 68M, 117M parameters, 16 experts, fp16), the final layout difference is
0.81, 0.88 and 0.86 of the seed level, and the share of near-tie decisions grows slightly with
size (6.6\%, 6.8\%, 7.8\%). The larger models diverge earlier: at step 32 their layout difference
is about 100$\times$ that of the base model (Appendix Table~\ref{tab:size} and
Figure~\ref{fig:size}). Within this range, size does not change the overall picture.

% =====================================================================
\FloatBarrier
\section{What This Means in Practice}
\label{sec:practice}

\paragraph{Do not expect the same MoE back from a different layout.} Even with the same seed,
data and batch, a new layout gives an MoE that is about 80\% as different as a new seed. If you
need the same model---to debug a loss spike, audit a result, or reproduce a release---keep the
layout fixed and use bitwise-deterministic training \citep{nvidia2026determinism}. Higher
precision alone does not help an MoE much; it helps a dense model a lot.

\paragraph{Compare methods on the same layout, with several seeds.} A change of layout alone
moves an MoE almost as much as a change of seed. A difference between two methods trained on
different layouts that is smaller than the seed-to-seed spread can simply be the layout.

\paragraph{During pretraining, avoid layout switches in the first part of the run.} For elastic
or preemptible pretraining jobs, a layout switch at 84\% of pretraining changed the model little,
while a switch at 10\% (just after warm-up) changed it a lot. If the cluster size must change,
doing it after the loss has flattened costs much less reproducibility than doing it right after
warm-up. We did not test layout switches during learning-rate decay, mid-training or
post-training.

\paragraph{Watch the router, not only the loss.} Loss did not reveal any of these differences.
The share of near-tie decisions and the share of routing decisions that change are cheap to log
and show how reproducible a run is. Expect them to get worse with more, smaller experts.

\paragraph{Plan for routing to be the fragile part.} For work that depends on specific
experts---routing replay in reinforcement learning \citep{ma2025r3}, expert pruning, or
interpretability---fine-grained MoEs need either deterministic, fixed-layout training or a
recorded routing table, because their expert assignments do not survive a layout change.

% =====================================================================
\section{Limitations}
\label{sec:limitations}

\textbf{Scale.} Our models have 25M--117M parameters and are trained on 20M tokens, far below
production budgets; the base model sees only a small fraction of a compute-optimal token budget.
The exact ratios (for example, 0.8$\times$ the seed level) will change with longer training.
\textbf{Hardware.} CPU runs use one node with up to 8 processes; GPU runs share a single T4.
Communication uses gloo, not NCCL, and we do not test tensor or pipeline parallelism.
Production training in bf16 or FP8 adds more rounding noise per step; based on our fp16 results
we expect the MoE effect to be at least as strong, but we have not measured it. \textbf{Repeats.}
The expert-count sweep and model-size study use one seed and one changed layout per setting.
\textbf{fp16 1-ULP control.} On the GPU, a 1-ULP change to an fp32 master weight was often lost
when the weight was cast to fp16, giving identical runs; we therefore use only seed and layout
comparisons for the GPU runs. \textbf{Quality.} We find no quality cost from per-machine rules at
our scale, but large-scale results show that per-micro-batch load balancing can hurt quality
\citep{qiu2025demons}; our result should not be read as saying these rules are harmless in
general. \textbf{Training phase.} Every run is pretraining from scratch with a constant learning rate
after warm-up, and the resume experiment uses only two switch points (10\% and 84\% of
pretraining). We make no claim about switching layouts during learning-rate decay or annealing,
mid-training, or post-training such as fine-tuning or reinforcement learning.
\textbf{Capacity.} We use a capacity factor of 1.25 with token dropping during training;
many production MoEs are dropless.

% =====================================================================
\section{Conclusion}
\label{sec:conclusion}

We asked whether the parallelism layout changes a trained Mixture-of-Experts model. It does not
change how good the model is, but it does change which model you get: about 80\% as much as a
new random seed, with most routing decisions different. The cause is the router: decisions where
two experts are nearly tied flip under the smallest rounding difference, and the runs then drift
apart on their own. Finer experts create more near-ties and make routing much less reproducible,
yet the outputs drift no further. When a pretraining run is resumed on a new layout, the point
at which it switches matters: a switch 10\% into pretraining changes the final model far more than
a switch at 84\%. A dense model of the same active size is far more stable and
depends on how much numerical noise there is, while the MoE diverges at any precision. For anyone
who needs the same MoE twice, the layout belongs in the experiment record, and fixed-layout
deterministic training is the only reliable way to get it.

\paragraph{Code and data.} The training code, the correctness checks, all run logs and the
measured data behind every table and figure will be released with this paper.

% =====================================================================
\FloatBarrier
\bibliography{references}

% =====================================================================
\appendix

\section{Correctness Checks and Implementation}
\label{app:checks}

Before training, each job ran a self-test with four processes on a tiny model. (1) For EP sizes
1, 2 and 4, with both balancing methods, the loss and the combined gradients of the multi-process
run matched a single-process run (relative error below $10^{-4}$ in fp32). (2) Three optimizer
steps run twice from the same start gave bit-identical weights. (3) A checkpoint saved under one
layout and loaded under another reproduced every weight and optimizer state exactly. (4) Every
per-machine variant ran without errors. Both the CPU and the GPU self-tests passed.

Expert-parallel communication is built only from sum-reductions. To gather rows from other
processes, each process writes its rows into a zero-filled buffer and the buffers are summed;
since adding zero is exact, gathering itself adds no rounding. All per-machine rules have a
``global'' version that gives the same result in every layout, so the numerical-only runs
change nothing but the order of additions. Checkpoints are written every 15 minutes in a
layout-independent format, which is what makes the resume experiments possible. Data
order is fixed per step: step $s$ always uses the same 32 sequences, and each process takes a
contiguous slice of them.

\section{Additional Results}
\label{app:extra}

\begin{table}[!htbp]
\centering
\includegraphics[width=\linewidth]{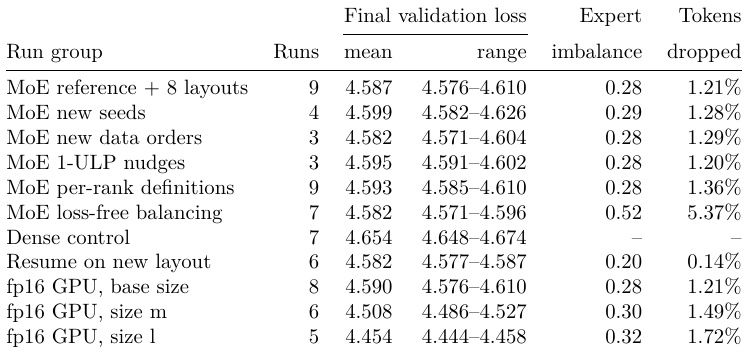}
\caption{Final validation loss and routing health by run group. ``Expert imbalance'' is the mean
over layers of the most-used expert's share divided by the average share, minus one.
``Tokens dropped'' is the share of routing assignments dropped by the capacity limit. Resumed runs
report statistics only after the switch point, where imbalance is lower.}
\label{tab:quality}
\end{table}

\begin{table}[!htbp]
\centering
\includegraphics[width=\linewidth]{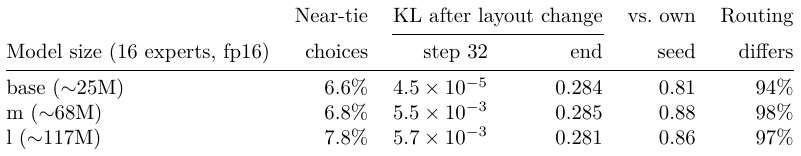}
\caption{Effect of model size (16 experts, fp16). The layout change is DP2/EP2.}
\label{tab:size}
\end{table}

\begin{figure}[!htbp]
\centering
\includegraphics[width=\linewidth]{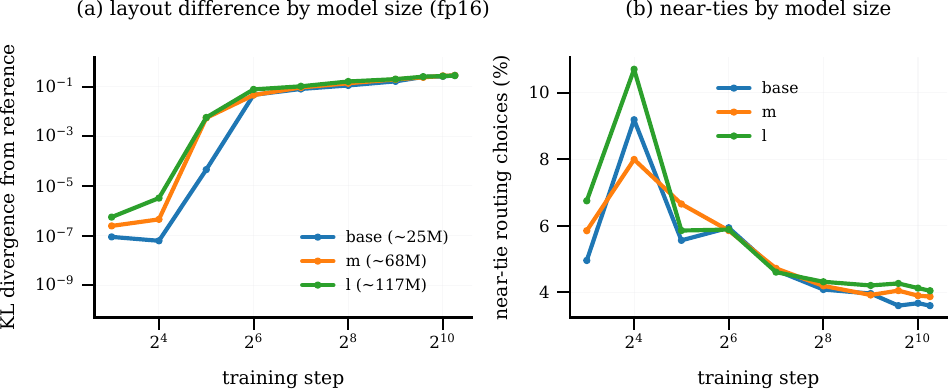}
\caption{(a) Difference after a layout change for three model sizes. (b) Share of near-tie routing
decisions during training.}
\label{fig:size}
\end{figure}

\FloatBarrier
\section{Reproducibility}
\label{app:repro}

All runs used the settings in Table~\ref{tab:setup}. CPU runs took about 2.2 hours each on one
16-vCPU node; a single-process base-size GPU run took about 13 minutes on one T4. Every number in this paper is read from
the CSV files in \texttt{data/}, and the scripts \texttt{figures/generate\_figures.py} and
\texttt{figures/make\_tables.py} regenerate every figure and table from them.

\end{document}